\documentclass[journal]{IEEEtran}
\usepackage{cite}
\usepackage{amsmath,amssymb,amsfonts}
\usepackage{algorithmic}
\usepackage{algorithm}
\usepackage{graphicx}
\usepackage{stmaryrd}
\usepackage{xcolor}
\usepackage{array}
\usepackage{commath}
\usepackage{sidecap}
\usepackage{stfloats}
\usepackage{tabularx, boldline}
\usepackage{rotating,booktabs,multirow}
\usepackage{mathtools}
\usepackage{flexisym}
\usepackage{breqn}
\usepackage{url}
\usepackage{adjustbox}
\usepackage{makecell}

\usepackage{textcomp}
\usepackage{amsmath}
\usepackage{txfonts}
\usepackage{graphicx}
\usepackage{acronym}
\usepackage{tabularx, boldline}
\usepackage{rotating,booktabs,multirow}
\usepackage{enumerate}
\usepackage{etoolbox}
\usepackage[]{nomencl}   
    \makenomenclature

\providetoggle{nomsort}
\settoggle{nomsort}{true} 

\makeatletter
\iftoggle{nomsort}{%
    \let\old@@@nomenclature=\@@@nomenclature        
        \newcounter{@nomcount} \setcounter{@nomcount}{0}%
        \renewcommand\the@nomcount{\two@digits{\value{@nomcount}}}
        \def\@@@nomenclature[#1]#2#3{
          \addtocounter{@nomcount}{1}%
        \def\@tempa{#2}\def\@tempb{#3}%
          \protected@write\@nomenclaturefile{}%
          {\string\nomenclatureentry{\the@nomcount\nom@verb\@tempa @[{\nom@verb\@tempa}]%
          \begingroup\nom@verb\@tempb\protect\nomeqref{\theequation}%
          |nompageref}{\thepage}}%
          \endgroup
          \@esphack}%
      }{}
\makeatother

\usepackage{authblk}
\def\BibTeX{{\rm B\kern-.05em{\sc i\kern-.025em b}\kern-.08em
    T\kern-.1667em\lower.7ex\hbox{E}\kern-.125emX}}

\ifCLASSINFOpdf

\else

\fi

\begin{document}


\title{Cardiotocography Signal Abnormality Detection based on Deep Unsupervised Models}

\author{Julien Bertieaux, Mohammadhadi~Shateri,~\IEEEmembership{Member,~IEEE,}
        Fabrice~Labeau,~\IEEEmembership{Senior Member,~IEEE}, Thierry~Dutoit,~\IEEEmembership{Member,~IEEE}
        
\thanks{J. Bertieaux and T. Dutoit are with the Department of Electrical Engineering, FPMs Umons University, Belgium.\\ Email:\{julien.bertieaux@student.umons.ac.be,thierry.dutoit@umons.ac.be\}}
\thanks{M. Shateri is with the Department of Systems Engineering, École de technologie supérieure, QC, Canada. (Email: mohammadhadi.shateri@etsmtl.ca)}
\thanks{F. Labeau are with the Department of Electrical and Computer Engineering, McGill University, QC, Canada. (Email: fabrice.labeau@mcgill.ca)}
}

\markboth{}{}
%

\maketitle


\begin{abstract}

Cardiotocography (CTG) is a key element when it comes to monitoring fetal well-being. Obstetricians use it to observe the fetal heart rate (FHR) and the uterine contraction (UC). The goal is to determine how the fetus reacts to the contraction and whether it is receiving adequate oxygen. If a problem occurs, the physician can then respond with an intervention. Unfortunately, the interpretation of CTGs is highly subjective and there is a low inter- and intra-observer agreement rate among practitioners. This can lead to unnecessary medical intervention that represents a risk for both the mother and the fetus. Recently, computer-assisted diagnosis techniques, especially based on artificial intelligence models (mostly supervised), have been proposed in the literature. But, many of these models lack generalization to unseen/test data samples due to overfitting. Moreover, the unsupervised models were applied to a very small portion of the CTG samples where the normal and abnormal classes are highly separable. In this work, deep unsupervised learning approaches, trained in a semi-supervised manner, are proposed for anomaly detection in CTG signals. The GANomaly framework, modified to capture the underlying distribution of data samples, is used as our main model and is applied to the CTU-UHB dataset. Unlike the recent studies, all the CTG data samples, without any specific preferences, are used in our work. The experimental results show that our modified GANomaly model outperforms state-of-the-arts. This study admit the superiority of the deep unsupervised models over the supervised ones in CTG abnormality detection.




\end{abstract}


\IEEEpeerreviewmaketitle

\section{Introduction}\label{intro}


\IEEEPARstart{C}{a}rdiotocography (CTG) is used throughout pregnancy to ensure that the fetus is receiving oxygen adequately. During labor, its purpose is to detect if the fetus is in distress in order to respond with an intervention such as a cesarean section. The fetal well-being is assessed by looking at the fetal heart rate (FHR) signal in response to the uterine contractions (UC) signal. Unfortunately, the interpretation of the results is highly subjective \cite{hruban2015goldenstandard} even when it is done based on internationally recognized guidelines such as those of FIGO (International Federation of Gynecologists and Obstetricians) \cite{ayres2015figo}. This has spurred efforts to develop computer-assisted interpretations that can be of great help for physicians by reducing the diagnosis time and uncertainty. Thus, there have been many studies on developing such a computer-assisted approach, especially the most recent ones that are based on artificial intelligence techniques.

The current state-of-the-art studies can be sorted according to the datasets and the methodologies applied to these datasets. 
Before discussing them via these aspects, two important points might be of interest to be noted. First, studies such as~\cite{georgieva2017cohortstudy,Ogasawara2021deepctg,petrozziello2019mcnn,petrozziello2018deepctg} used their own private datasets and although include enough CTG data samples for training a deep model, those datasets are not publicly available for the sake of regenerating the results. Second, depending on the way that a CTG sample is labeled as normal or abnormal, the results of these works might be completely different. Therefore, these studies and their findings should be taken with caution since they are not directly comparable. 

With regards to the dataset, in studies such as~\cite{subasi2020classificationbagging, sahin2015classificationml , nagendra2017ctgsvmrandomforest} CTG datasets include features extracted from the raw temporal CTG signals are used. As an example of such datasets, the UCI Machine Learning Repository Cardiotocography dataset~\cite{bache2010cardiotocography} can be named that only compiles 23 features based on the FIGO guidelines \cite{bache2010cardiotocography , ayres2000sysporto}. Although using these datasets might reduce the difficulties in training a deep AI model, these features are just a synthesis of temporal signals and therefore the AI model is not given the opportunity to detect other relevant features. 
In our work, we use the CTU-UHB dataset proposed in~\cite{chudavcek2014open} that not only is publicly available, but also includes temporal CTG signals and has been used extensively in the literature~\cite{alyousif2021survey}. Table~\ref{tab:recent_works} lists some of the recent studies who used this dataset in their work.

\begin{table}[htbp]
	\centering
	\caption{Some recent studies that used CTU-UHB dataset (552 data samples).}
	\begin{adjustbox}{width=0.47\textwidth}
		\begin{tabular}{l c c}
			\toprule
            \textbf{Study}  & \textbf{\makecell{Labelling criteria\\ (pathological)}} & \textbf{Model}\\
            \midrule[0.1pt]
            
            \textbf{Ajirak et al. (2022) \cite{ajirak2022boost}} & \makecell{pH $<$ 7.05} & 
            \makecell{Decision Tree, Random Forest,\\ and  Support Vector Machine}\\
            \midrule[0.1pt]
            \textbf{Liu et al. (2021) \cite{liu2021attention}} &  \makecell{pH $\leq$ 7.15} & \text{CNN-BiLSTM + Attention, DWT} \\
            \midrule[0.1pt]
            \textbf{Fergus et al. (2021) \cite{fergus2021ctgcnn}} &  \makecell{pH $<$ 7.20 } & \text{1-D CNN} \\
            
            \midrule[0.1pt]
            
            \textbf{Baghel et al. (2021) \cite{baghel2022fhrnet}}&\makecell{pH $<$ 7.15}&\text{1-D FHRNet (1-D CNN)}\\
            \midrule[0.1pt]
            
            \textbf{Zhao et al. (2019) \cite{zhao2019deepfhr}}&\makecell{pH $<$ 7.15}&\text{DeepFHR (CNN)}\\
			\bottomrule
		\end{tabular}
	\end{adjustbox}
	\label{tab:recent_works}
\end{table}

In terms of the methodologies, we can name supervised and unsupervised approaches where the former is the dominant approach used in the literature~\cite{alyousif2021survey}. For example, in~\cite{ajirak2022boost} the classical machine learning models such as decision tree, support vector machine, and ensemble learning classifiers were used. More sophisticated deep learning methods, e.g. those that incorporate Convolutional Neural Network (CNN), long short-term memory (LSTM), or attention models, have also been considered in the literature~\cite{fergus2021ctgcnn,zhao2019deepfhr,Ogasawara2021deepctg,liu2021attention,baghel2022fhrnet}. However, the validity of the available models can be undermined due to two main issues. First of all, the performance of some of them was evaluated on the training and validation data samples. This makes the models lack generalization, i.e. not being effective on the test (unseen/untouched) data samples. It should be noted that validation data samples are used to set the hyperparameters of a model and once it is done, the actual performance of the model should be evaluated on a test dataset that remained untouched by the model. Secondly, many of the papers used accuracy metrics to evaluate their models\cite{alyousif2021survey}. Unfortunately, accuracy is not a reliable metric for our problem because most CTG datasets, e.g. CTU-UHB dataset, are imbalanced and thus using accuracy as a measure of evaluation can be misleading. For instance, if the anomalous class represents only $10\%$ of the data samples, then a no-skill (random) classifier who, regardless of the data samples, decides all the time in favor of the major class can reach an accuracy $90\%$.  


The unsupervised clustering approach was used for the first time by Przybyła et al. in~\cite{przybyla2009unsupervised} where fuzzy c-means clustering was applied to FHR signals. But they concluded that the unsupervised clustering can not be used as a reliable approach to assess the FHR state. Another unsupervised clustering, based on the deep Gaussian Processes (deep GPs)~\cite{damianou2013deep}, was studied in~\cite{feng2018supervised}. Although this approach was applied to the CTU-UHB dataset (that is of our interest), only 10 FHR recordings (3 from the abnormal class and 7 from the normal class) were considered. In another study published recently in~\cite{yang2022unsupervised} a density-based spatial clustering method called DBSCAN was considered. In that study, first, the UC-dependent FHR segments were determined and mapped to a lower dimension using the Gaussian process latent variable model and then the DBSCAN model was used for clustering. This approach was just applied to a portion of data samples in CTU-UHB dataset where the confidence about their class labels was more than $50\%$ (83 normal and 50 abnormal samples). An unsupervised anomaly detection method using phase space reconstruction of FHR segments was proposed in~\cite{yang2021unsupervised}. But the model was applied just to either synthetic CTG data samples or some selected CTGs in CTU-UHB dataset ( the five samples with lowest pH values and 80 highly confident normal samples). This suggests that these unsupervised models proposed in the literature are just capable to be applied to those data samples that are confidently normal or abnormal. Overall, the existing literature on unsupervised learning  is sparse, and existing studies lack in terms of the quality of the testing data and of the performance metrics used.

 In this work, we adopt an unsupervised learning approach where, different from the literature, we use deep unsupervised models trained in a semi-supervised manner for anomaly detection in CTG signals. In our study, we consider the whole data samples in the CTU-UHB dataset (not just some selected data samples as done in the literature). We start with the Isolation Forest as a classical unsupervised model for anomaly detection. Then we incorporate the use of more sophisticated models such as the deep Autoencoder. We use the GANomaly model, with a simple modification in its loss function for the sake of capturing the underlying distribution in the data, as our main model for abnormality detection in CTG signals. To evaluate the performance of the proposed models, we use metrics such as F1-score, balanced accuracy, and precision-recall curve that are more appropriate for imbalanced datasets. In addition, we make sure that the data is divided into three parts including train, validation, and test. The train and validation are used respectively to train the model and set the hyperparameters. The test data samples remained untouched by the model until the last stage and is used only for the final evaluation. This study proposes a clear path for the development and evaluation of the models for abnormal CTG signal detection and emphasizes the superiority of the deep unsupervised approaches compared with the state-of-the-art supervised models used in the literature~\cite{ajirak2022boost, liu2021attention, georgoulas2017investigatingph,subasi2020classificationbagging,sahin2015classificationml}. The contributions of this manuscript can be summarized as follows:
 \begin{itemize}
     \item [-] We propose a modified GANomaly model trained in a semi-supervised manner to detect abnormal fetal heart rate samples. To the best of our knowledge, this is the first work that propose a semi-unsupervised anomaly detection based on deep neural networks for abnormality detection in CTG signals.
     \item [-] The proposed model is applied to CTU-UHB that is a real-world dataset and unlike the state-of-the-arts, all the data samples within this dataset are used.
     \item [-] The process of training, hyperparameter tuning, and testing are done carefully to prevent overfitting and to ensure convergence of the proposed model. According to the experimental results, our modified GANomaly model shows to outperform the state-of-the-arts including classical machine learning models and deep classifiers.
     \item [-] The results of this work clearly emphasize on the superiority of the unsupervised (and semi-supervised) models to the supervised ones for the CTG abnormality detection. Therefore this study suggests a new path (i.e. unsupervised abnormality detection) in addition to the supervised approach in this field.\\
 \end{itemize}


The rest of the paper is organized as follows. In Section~\ref{dataprep} an in-depth discussion about the dataset used in this study is presented. Section~\ref{sec:back_motiv} presents the the baseline models (used in recent studies), our proposed models in details, and the evaluation metrics used in our work. The details regarding experiments and extensive results from these experiments are presented and discussed in Section \ref{sec:results}. Finally, some concluding remarks are presented in Section \ref{sec:conclusion}.


\section{Data} \label{dataprep}
In this study, the CTU-UHB dataset is used. This is an open-access dataset proposed in~\cite{chudavcek2014open} and includes 552 intrapartum recordings. These recordings were selected from 9164 CTGs recorded between 2010 and 2012 at the University Hospital of Brno, Czech Republic. The selection criteria in this dataset include gestational age over 36 weeks, no known developmental defects, and singleton pregnancy. Such selection criteria enabled the creation of a homogeneous group of patients for the dataset. This was necessary because, for instance, the FHR signal tends to change with gestational age \cite{serra_bellver_2009}.
Moreover, it was chosen to work only with CTGs that are provided with biochemical parameters of the umbilical arterial blood sample. The analysis of the blood sample contains the pH which will be used to label the samples.

This dataset consists of two kinds of information regarding cardiotocography measurements such as the fetal heart rate (or FHR) and the uterine contractions (or UC) in parallel. Due to the lack of quality of the UC signal, similar to most of the state-of-art papers we only used the FHR signal~\cite{alyousif2021survey}. But it is worth to be noted that in order to make their diagnosis, doctors observe how the FHR reacts after contractions. It is therefore important to consider using both signals in future works. The CTU-UHB dataset also contains a series of additional information such as the analysis of umbilical artery blood sample (i.e. pH, pCO2, pO2, BDecf) or the Apgar1/5 score. This information allowed us to split the recordings into 2 groups (normal and abnormal). Among the aforementioned parameters, pH is the one used mostly to label the CTG signals. But indeed, the pH alone is not sufficient for an effective discrimination and thus, both pH and Apgar score (after 1 minute or Apgar1) are used. This Apgar1 is a subjective metric about the well-being of the baby 1 minute after the birth. Regarding the threshold values applied to the pH and Apgar1, there is no general agreement in the literature~\cite{chudavcek2014open}. In this work, following the studies~\cite{Ogasawara2021deepctg, fergus2021ctgcnn} we consider recordings with pH $<7.20$ and Apgar1 score $<7$ as abnormal. Those thresholds define an abnormal signal as a suspicious CTG that might need the obstetrician's attention. By labeling the CTG signals this way, we obtain 182 abnormal and 370 normal recordings. It should be noted that the labeling process could be enhanced by using more of the information available, such as BDef (base deficit). The base deficit can indicate the severity of acidosis while the pH is the main indicator of its presence. In the scope of this work, we did not need the base deficit, but it could be considered in future works.



\begin{figure}[htbp!]
    \centering
    \includegraphics[width=1\linewidth]{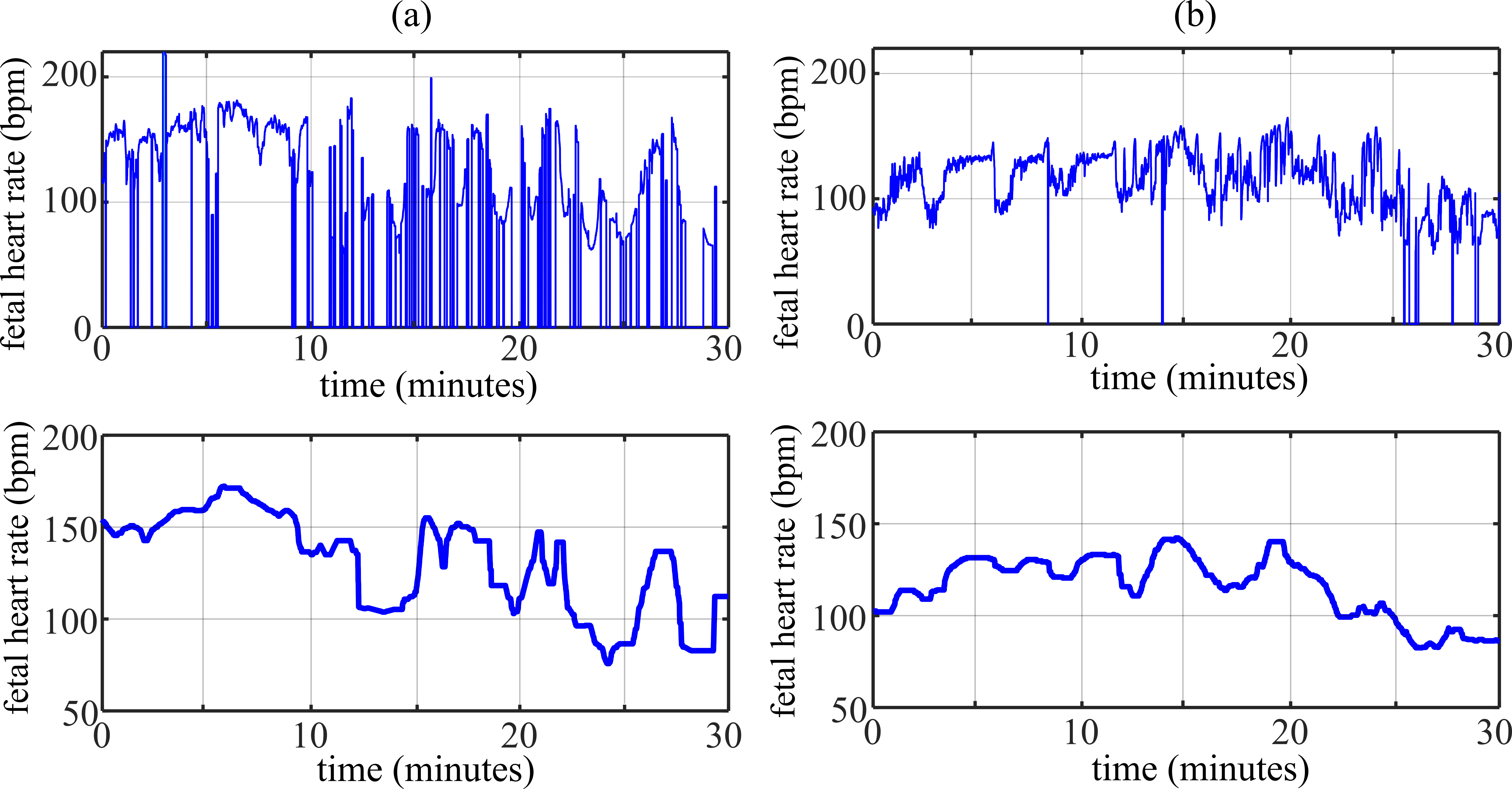}
    \caption{An example of the fetal heart rate signal before and after preprocessing for (a) abnormal (suspicious) class and (b) normal class.}
    \label{fig:prepro}
\end{figure}

Regarding the preprocessing of the proposed data samples, three major detrimental effects needed to be dealt with including (i) missing values, (ii) Gaussian/burst noise, and (iii) the mismatch between maternal heart rate (MHR) and fetal heart rate. 
First, FHR values greater than 200 bpm or less than 50 bpm that might exist due to several reasons during the test e.g., movement of mother/fetal or the ultrasound probe, are removed and treated as missing values~\cite{cesarelli2007algorithm}. Then linear interpolation is applied to deal with all the missing values\cite{liu2021attention}. Finally, a median filter is used to smooth the FHR signals~\cite{liang2021automatic}. Fig.~\ref{fig:prepro} represents an example of the FHR signal before and after preprocessing. It worth to be noted that where ever it is possible, it is mandatory to check the maternal pulse to avoid erroneous recording of maternal heart rate as fetal. Unfortunately, we could not discriminate maternal and fetal heart rates with quantitative criteria. In fact, the sensors used to detect the FHR signal sometimes measure the MHR signal. To compensate for this, physicians also check the MHR to avoid any erroneous association \cite{alyousif2021survey}. In the CTU-UHB dataset, the MHR signal is not available so we could not use it to know if the signal measured is truly the FHR.

\section{Methods}\label{sec:back_motiv}
It was discussed in the introduction section that there are two main approaches including supervised and unsupervised techniques for abnormality inference in CTG signals. In our study, the (supervised) classification models used in the recent studies (based on both classical machine learning and deep learning models) are used and compared with our unsupervised anomaly detection models. During this work, we assume that the dataset $\{(x_i,y_i)\}_{i=1}^{N}$ includes $N$ data samples of CTG signals, e.g. fetal heart rate, in which $x_i\in\mathcal{R}^d$ is a d-dimensional feature vector with an associated label $y_i\in\{0,1\}$ where labels 0 and 1 respectively refer to normal and abnormal classes. In addition, it is assumed that there are $n$ data samples from the normal class where $n>N-n$, i.e. the dataset is imbalanced.

\subsection{Baselines: Classification approach}
In the classification approaches, the goal is to develop a mapping $f:\mathcal{R}^d\longrightarrow\{0,1\}$ that maps each feature vector $x_i$ to its associated label $y_i$. This can be done by minimizing the classification error. There are a wide range of classification models, from the family of classical machine learning models such as logistic regression, decision tree, support vector machine, and so on to those models based on deep neural networks and deep convolutional neural networks. However, we believe that the classification models based on deep learning might not be a good choice. The main reason is the lack of enough data samples in the available real-world datasets while the deep classifiers are data hungry. More specifically, applying deep classification models in this case can result in overfitting which means that although the model might work well during the training, it lacks generalization on the test (unseen) data samples. In this work, to compare the performance of our presented models with those in the literature on the same dataset, following the study~\cite{ajirak2022boost}, two well-known models from the classical machine learning family called Random Forest and Support Vector Machine, both with balanced class weights, are used. In addition, a more sophisticated classifier proposed in~\cite{liu2021attention} based on deep CNN, BiLSTM, and Attention network is used where features based on discrete wavelet transform (DWT) are incorporated. For information regarding the details of these models, readers are referred to the references~\cite{ajirak2022boost,liu2021attention}.

\subsection{Main models: Unsupervised anomaly detection approach \label{unsupervisedanomaly}}
The supervised learning models require the availability of enough labelled data samples. In many practical cases, including this work, there is a lack of enough data samples, specially samples from a minority class of our interest (i.e., abnormal class). Although some techniques such as weighting approaches (we mentioned for Random Forest and Support Vector Machine in the classification section)~\cite{fernandez2018learning} or data augmentation, e.g. by using generative adversarial networks, might help, in some cases one can benefit more from unsupervised models such as those based on anomaly detection techniques. An anomaly, also referred to as an outlier, is a data sample that deviates considerably from the rest of data samples. There are different techniques proposed in the literature, especially those based on deep learning, and used to detect anomalies in different applications, e.g. intrusion detection, fraud detection, industrial damage detection, public health, and so on. For a survey, the readers are referred to~\cite{xia2022gan,ruff2021unifying}. In this work, we adopt the unsupervised anomaly detection approaches starting from Isolation Forest as a classical approach and turning into our main model that is a modified version of a GANomaly, a sophisticated deep learning approach. 
\subsubsection{Isolation Forest} 
\begin{figure}[htbp!]
    \centering
    \includegraphics[width=1\linewidth]{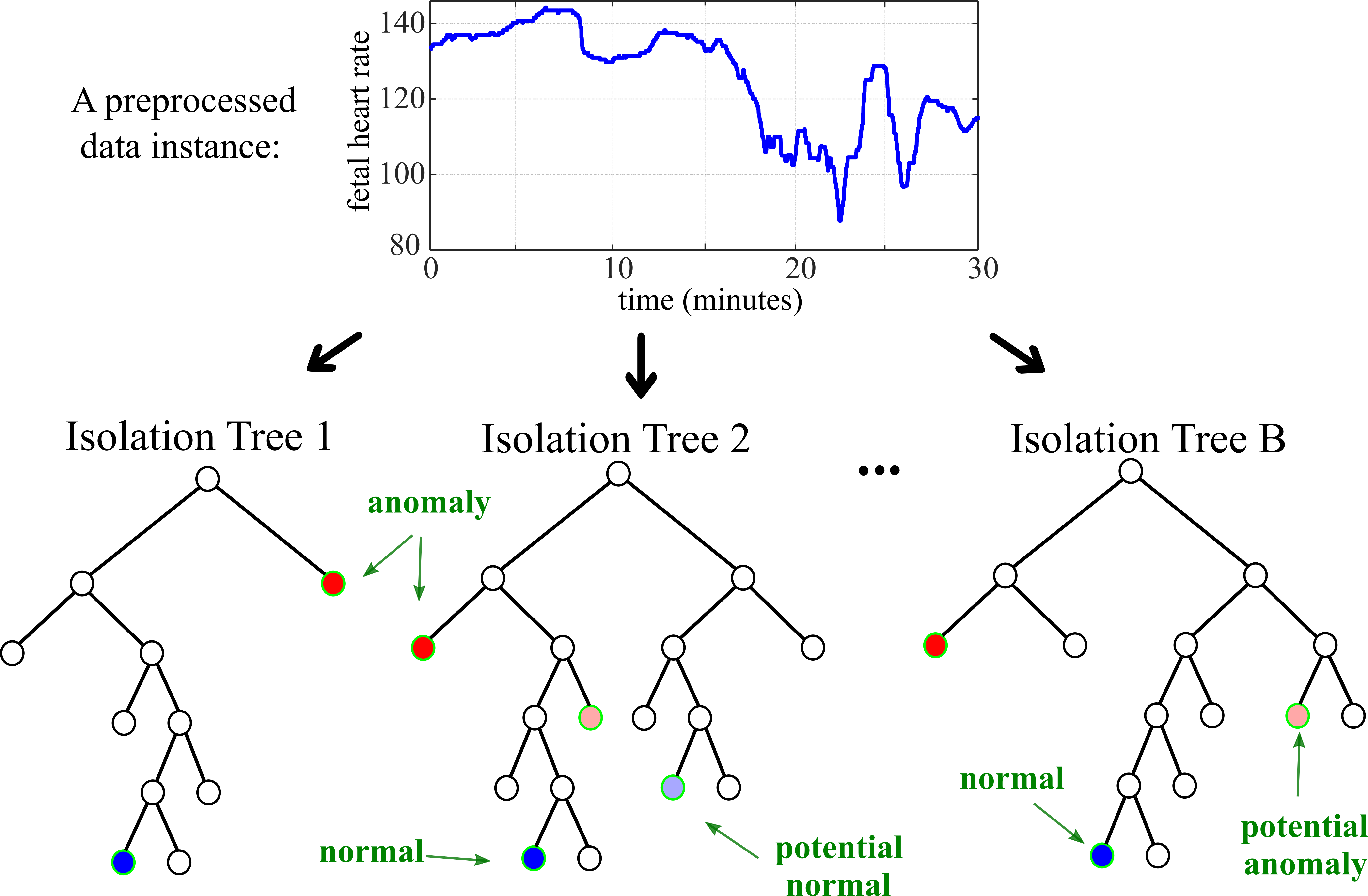}
    \caption{Isolation Forest anomaly detection based on an ensemble of $B$ isolation trees. The colored nodes refer to the isolation of data samples.}
    \label{fig:IF}
\end{figure}

Isolation Forest (IF) is an unsupervised anomaly detection technique proposed by Liu et. al.~\cite{liu2008isolation} and similar to Random Forest, IF is built based on Decision Trees. The main idea of IF technique is the fact that anomalies are few and different. Therefore, those data samples that go deeper into the trees, i.e. need more cuts to be isolated, are less likely to be anomalies. In other words, the data samples that trees found simpler to distinguish from the others (and so are found on the shorter branches of the trees) are more likely to be anomalies. Fig.~\ref{fig:IF} shows an IF based on an ensemble of $B$ isolation trees from which to train each tree: first a random sub-sample of the data is selected and then branching is done based on a random threshold applied to a randomly selected feature. This process continues until isolating each data sample (or reaching to the maximum depth of the tree). 
After creating an IF, each data sample in the test dataset is passed through all the trained isolation trees and then an anomaly score is assigned to it, depending on an aggregation of the depths required by each tree to isolate that data sample. \\

\subsubsection{Autoencoder}
\begin{figure}[htbp!]
    \centering
    \includegraphics[width=0.7\linewidth]{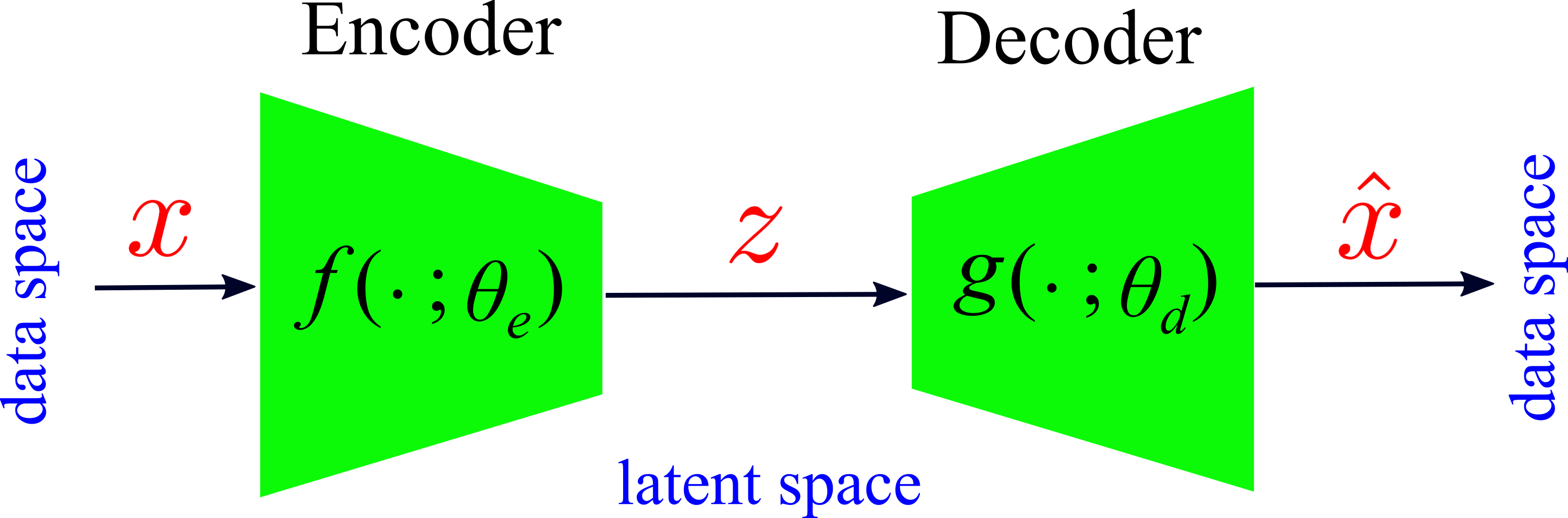}
    \caption{Framework of autoencoder. $\theta_{e}$ and $\theta_{d}$ are the learning parameters of the encoder and decoder, respectively.}
    \label{fig:AUE}
\end{figure}

The autoencoder (AUE), proposed by LeCun~\cite{lecun1987modeles}, is an unsupervised learning model mainly used to learn a low-dimensional representation of data in a non-linear approach. The framework of an AUE (see Fig.~\ref{fig:AUE}) includes an encoder and a decoder where both are modelled as deep neural networks. The encoder maps the data sample $x$ in the data space to a lower-dimension feature $z = f(x;\theta_{e})$ in the feature space (latent space) while the decoder reconstructs the data sample as $\hat{x} = g(z;\theta_{d})$ where $\theta_{e}$ and $\theta_{d}$  respectively refer to the learning parameters of the encoder and decoder.
The AUE framework is trained by minimizing the error associated with reconstructing the original data sample $x$ out of the latent $z$. Thus, the following optimization problem is used for the AUE:
\begin{equation} \label{eq:AUE_op} \underset{f,\; g}{\text{min}}  \;\mathbb{E}\left[|| X - \hat{X}||_1\right], \end{equation}
where $||.||_1$ refers to the $\ell_1$ norm and the expectation is in terms of $X$. It is important to be noted that the mapping that an AUE learns is very specific to the training data distribution and so if it is applied to an unseen data sample significantly different from the training set, a large reconstruction error should be expected. This is the main motivation for using an AUE as anomaly detection. More specifically, to use the AUE for anomaly detection, a semi-supervised approach is used where the AUE framework is trained on the normal data samples. So the AUE can learn the patterns available in the class of normal data samples and reconstruct them from a low-dimensional feature vector with a small error. Considering the optimization problem~\eqref{eq:AUE_op} an empirical loss function~\eqref{eq:AUE_loss} is used:
\begin{equation}\label{eq:AUE_loss}
    \mathcal{L}_{AUE}(\theta_{e},\theta_{d}) \coloneqq \frac{1}{n}\sum_{i=1}^{n} || x_i - \hat{x}_i||_1
\end{equation}
where $n$ is the number of normal data samples. At the test time, to determine whether a given data sample $x$ is an anomaly or not, an anomaly score is assigned based on the reconstruction error and the samples is flagged as an anomaly if its associated score is more than a specified threshold (see equation~\eqref{eq:aue_anomaly_score}).
\begin{equation} \label{eq:aue_anomaly_score}
    L_x \coloneqq \left\{
\begin{array}{lr}
	0\; (Normal) \hspace{0.9cm} s_x\leq\tau\\
	\\
	1\; (Abnormal)\; \hspace{0.5cm} s_x>\tau\\
\end{array}
\right.
\;\hspace{0.25cm}\;, s_x = ||x - g(f(x))||_{1}
\end{equation}
where $L_x$ and $s_x$ are the anomaly label and score assigned to data sample $x$, respectively. In addition, the parameter $\tau$ is a threshold that separates the abnormal class from the normal class. The threshold $\tau$ is considered as a hyperparameter and is determined based on validation data samples. \\

\subsubsection{Modified GANomaly} 
\begin{figure}[htbp!]
    \centering
    \includegraphics[width=0.95\linewidth]{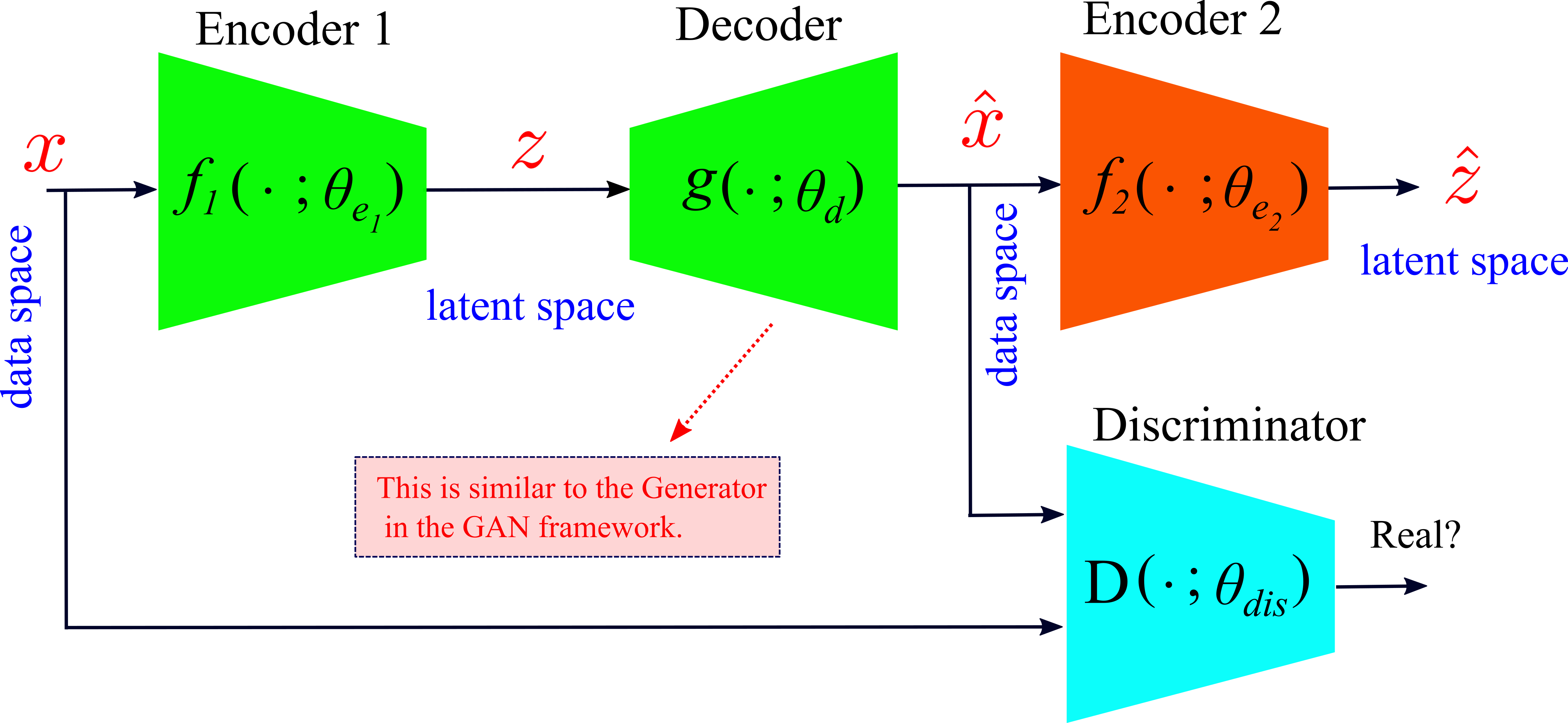}
    \caption{GANomaly model. $\theta_{e1}$, $\theta_{d}$, $\theta_{e2}$, and $\theta_{dis}$ are the learning parameters of the encoder 1, decoder, encoder 2, and the discriminator respectively.}
    \label{fig:Ganomaly_framework}
\end{figure}

A more sophisticated approach to detect anomalies, rather than AUE, is GANomaly model\cite{akcay2018ganomaly}. The main idea with the GANomaly model is to keep not only the reconstruction error minimized, but also capture the distribution of the normal class. To this end, the generative adversarial network (GAN) is incorporated into the AUE framework (see Fig.~\ref{fig:Ganomaly_framework}). The decoder in this framework has the same role as the generator in GAN framework where it generates data sample $\hat{x}$ from the latent $z$ by fooling the discriminator (adversarial learning) to believe that $\hat{x}$ comes from the same distribution as $x$. More details regarding GAN and its applications can be found in~\cite{goodfellow2014generative,9129799}.
The formulation of the GANomaly model can be stated based on the following minimax game:
\begin{align}\label{eq:GANomali_minimax}
    \underset{f_1,\; g,\; f_2}{min}\underset{D}{max}\; V(f_1,g,f_2,D) &=\lambda_{c}\mathbb{E}\left[||x - g(f_1(x)) ||_1\right] +\\ 
    &\lambda_{e}\mathbb{E}\left[||f_1(x) - f_2(g(f_1(x))) ||_1\right] +\nonumber\\
    &\lambda_{a}\left(\mathbb{E}\left[\log D(x)\right] +\mathbb{E}\left[\log(1-D(g(z))\right]\right)\nonumber
\end{align}
where $\lambda_{c}$, $\lambda_{e}$, and $\lambda_{a}$ are the weights determine the importance of each terms. These weights are considered as hyperparameters and their values are found based on validation data samples. The formulation of the GANomaly model includes three important terms each are assigned for a specific goal. The first term (also named as contextual loss) ensures that the original sample and its reconstructed version are close to each other contextually and the second term (also named as latent/encoder loss) seeks the same goal but in the latent space. For the third term, i.e. adversary loss, unlike the formulation in~\cite{akcay2018ganomaly}, we use the original loss function of GAN. This loss function that is similar to the cross entropy loss function between the training and generated data samples, is the appropriate way to learn the distribution of the training data samples. In other words, minimizing this loss can lead to having generated samples $\hat{x}$ coming from the same distribution as data samples $x$~\cite{goodfellow2014generative}. It should be noted that the adversary loss function used in~\cite{akcay2018ganomaly} is very similar to the latent loss (second term in~\eqref{eq:GANomali_minimax}) as both are defined in terms of features extracted from the data samples $x$ and $\hat{x}$. Thus, to prevent this redundancy we modified the loss function of the GANomaly in the way presented in equation~\eqref{eq:GANomali_minimax}. Training algorithm of the proposed GANomaly is presented in the Algorithm~\ref{AL_GANomaly}.
\begin{algorithm}
    \footnotesize
    \algsetup{linenosize=\tiny}
	\caption{Anomaly detection based on the modified GANomaly model. Batch size $B$, weights $\lambda_{c}$, $\lambda_{e}$, $\lambda_{a}$ and number of steps $k_{g}$, and $k_{d}$ are hyperparameters.}
	\label{AL_GANomaly}
	\begin{algorithmic}[1]
	\FOR {number of training iterations}
	    \FOR {$k_{d}$ steps}
	    \STATE Sample minibatch of $B$ examples $\{ x_{1},\dots, x_{B}\}$ from training set and find their associated latent $\{ z_{1},\dots, z_{B}\}$.
		\STATE Sample minibatch of $B$ examples $\{ x_{1},\dots, x_{B}\}$ from training set.
		\STATE Compute the gradient of $\mathcal{L}_{D}(\theta_{dis})$ and update $\theta_{dis}$ by applying the Adam optimizer~\cite{kingma2014adam}.
		\vspace{5pt}
		
		$\mathcal{L}_{D}(\theta_{dis}) \coloneqq  - \frac{1}{B} \sum_{i=1}^{B}\left[ \log D(x_i) + \log(1-D(g(z_i)))\right]$
        
		\ENDFOR
		
		\FOR {$k_{g}$ steps}
		\STATE Sample minibatch of $B$ examples $\{ x_{1},\dots, x_{B}\}$ from training set and find their associated latent $\{ z_{1},\dots, z_{B}\}$.
		\STATE Sample minibatch of $B$ examples $\{ x_{1},\dots, x_{B}\}$ from training set.
		\STATE Compute the gradient of $\mathcal{L}_G(\theta_{e_1},\theta_{d},\theta_{e_2},\theta_{dis})$ and update $\theta_{e_1},\theta_{d},\theta_{e_2}$ by applying the Adam optimizer.
		\vspace{5pt}

       $\mathcal{L}_G(\theta_{e_1},\theta_{d},\theta_{e_2},\theta_{dis}) \coloneqq  \frac{1}{B}\sum_{i=1}^{B} \Big[\lambda_{c}||x_i - g(f_1(x_i)) ||_1 +$\\ 
       $\lambda_{e}||f_1(x_i) - f_2(g(f_1(x_i)))||_1+-\lambda_{a}\log D(g(z_i))\Big]\nonumber$
       
       \ENDFOR
       
       \vspace{5pt}
       
		\ENDFOR
	\end{algorithmic}
\end{algorithm}
After training the GANomaly based on the Algorithm~\ref{AL_GANomaly}, similar approach as the one mentioned for AUE in equation~\eqref{eq:aue_anomaly_score} is used to assign anomaly label and score to the test data samples.

\subsection{Performance Metrics}
It was mentioned that in this work the goal is to develop models to detect abnormal class which here is referred to suspicious case of CTG signals. Considering the suspicious class as the class 1 or positive class and so the normal CTG signal as class 0 or negative class, after applying the proposed supervised and unsupervised anomaly detection approaches (discussed in the previous subsections) two types of error can be assumed. The first type of error is associated with a false positive (FP) which in our case means a false suspicious while the second type of error is a false negative (FN). The performance of any anomaly detection model can be evaluated in terms of minimizing these two types of error. This usually is done by maximizing the accuracy defined as follows:
\begin{equation*}\label{eq:acc}
    \text{accuracy} = \frac{TP+TN}{TP + TN + FP + FN}
\end{equation*}
where TP and TN are true positive and true negative, respectively. For the cases where the dataset is imbalanced, accuracy could be misleading. The reason is due to this fact that a no-skill model who decides in favor of the major class regardless of the input would have a high accuracy. To prevent this, other measures are presented below\cite{saito2015precision}:
\begin{align}\label{eq:metrics}
    &\text{Balanced accuracy} = 0.5\times\left[\frac{TP}{TP+FN} + \frac{TN}{TN+FP}\right] \nonumber\\
    &\text{Precision} = \frac{TP}{TP + FP} \nonumber\\
    &\text{Recall} = \frac{TP}{TP + FN} \\
    &\text{F1-score} = 2\times\frac{\text{Precision}\times \text{Recall}}{\text{Precision}+ \text{Recall}}\nonumber
\end{align}
The metrics in equation~\eqref{eq:metrics} will be used in this study to evaluate the performance of the proposed models.

\section{Results and Discussion}\label{sec:results}
After preprocessing the data (explained in the section~\ref{dataprep}), the data samples are split into two disjoint parts including training and test datasets with ratio $90-10$. Thus, we come up with $56$ data samples in the test data set (includes $19$ abnormal samples) and $496$ data samples in the training set (with $163$ abnormal samples). It was discussed that each anomaly detection models (either supervised or unsupervised) has several hyperparameters, i.e. parameters that cannot be learned directly by the model. To set the values of these hyperparameters, we need to use a portion of the training dataset as a validation dataset. The values of hyperparameters are set by the grid search to obtain the best performance (in this study, the best F1-score) on the validation data samples. After setting the values of hyperparameters, the ultimate performance of models are determined using the test dataset (unseen/untouched data samples). 

In this study, for all the models except GANomaly, $10\%$ of the training is used as the validation. For the GANomaly model however, $40\%$ of the training set is used as validation dataset and to keep enough data samples for training phase, re-sampling of the training data samples is applied. For the Random Forest model, a total number of $10$ decision trees with maximum depth $2$ are used. Table~\ref{tab:hyperparameters} presents a list of the values of hyperparameters and model architectures used in our work.
\begin{table}[htbp]
	\centering
	\caption{List of hyperparameters and model architectures used in this work. These are set to have the best F1-score on the validation data samples.}
	\begin{adjustbox}{width=0.48\textwidth}
		\begin{tabular}{l c }
			\toprule
            \textbf{Model} & \textbf{Hyperparameters} \\
            \midrule[0.1pt]

            \textbf{CNN-BiLSTM + Attention, DWT~\cite{liu2021attention}}& \makecell{the same structure as proposed in~\cite{liu2021attention} is used.}\\
            \midrule[0.1pt]
            
            \textbf{Balanced Random Forest~\cite{ajirak2022boost}}& \makecell{$10$ decision trees with maximum depth $2$}\\
            \midrule[0.1pt]
            \textbf{Support Vector Machine~\cite{ajirak2022boost}}& \makecell{radial basis function (rbf) kernels with parameter $C = 0.1$}\\
            \midrule[0.1pt]
            \textbf{Isolation Forest}& \makecell{$100$ isolation trees with contamination value $0.33$}\\
            \midrule[0.1pt]
            
            \multirow{3}{*}\text{\textbf{Autoencoder}}&\textbf{Encoder:} \makecell{three dense layers with relu activation function \\and 128, 64, and 16 units, respectively.} \\ \cmidrule(lr){2-2}
            & \textbf{Decoder:} \makecell{three dense layers with relu activation function \\and 16, 64, and 128 units, respectively.}\\ \cmidrule(lr){2-2}
            & \textbf{Optimizer:} \makecell{Adam with learning rate 0.001 and $\beta_1 = 0.99$\\
            with $\tau = \mu + \sigma$}\\
            
            \midrule[0.1pt]
            
            \multirow{4}{*}\text{\textbf{GANomaly}}&\textbf{Encoder 1$\&$2:} \makecell{three dense layers with leaky relu* activation\\ function and 128, 64, and 16 units, respectively.} \\ \cmidrule(lr){2-2}
            & \textbf{Decoder:} \makecell{three dense layers with leaky relu* activation\\ function and 16, 64, and 128 units, respectively.}\\ \cmidrule(lr){2-2}
            & \textbf{Discriminator:} \makecell{three dense layers with leaky relu* activation\\ function and a sigmoid activation for the\\ output layer,and number of units 128, 16,\\ and 1 , respectively.}\\ \cmidrule(lr){2-2}
            & \textbf{Optimizer:} \makecell{Adam with learning rate 0.0002 and $\beta_1 = 0.50$.\\ $\lambda_c$, $\lambda_e$, $\lambda_a$ with values 50, 1, and 1, respectively. \\
            $k_d = 1$ and $k_g = 2$ with $\tau = \mu + 5\sigma$}\\
			\bottomrule
			\multicolumn{2}{l}{*Leaky relu with parameter $\alpha = 0.2$} \\
			\multicolumn{2}{l}{**$\mu$ and $\sigma$ are the mean and standard deviation of training anomaly scores. } \\
		\end{tabular}
	\end{adjustbox}
	\label{tab:hyperparameters}
\end{table}

Considering the hyperparameters and model architectures presented in the Table~\ref{tab:hyperparameters}, we evaluate the performance of each model on the test data samples. 

\begin{figure}[htbp!]
    \centering
    \includegraphics[width=1\linewidth]{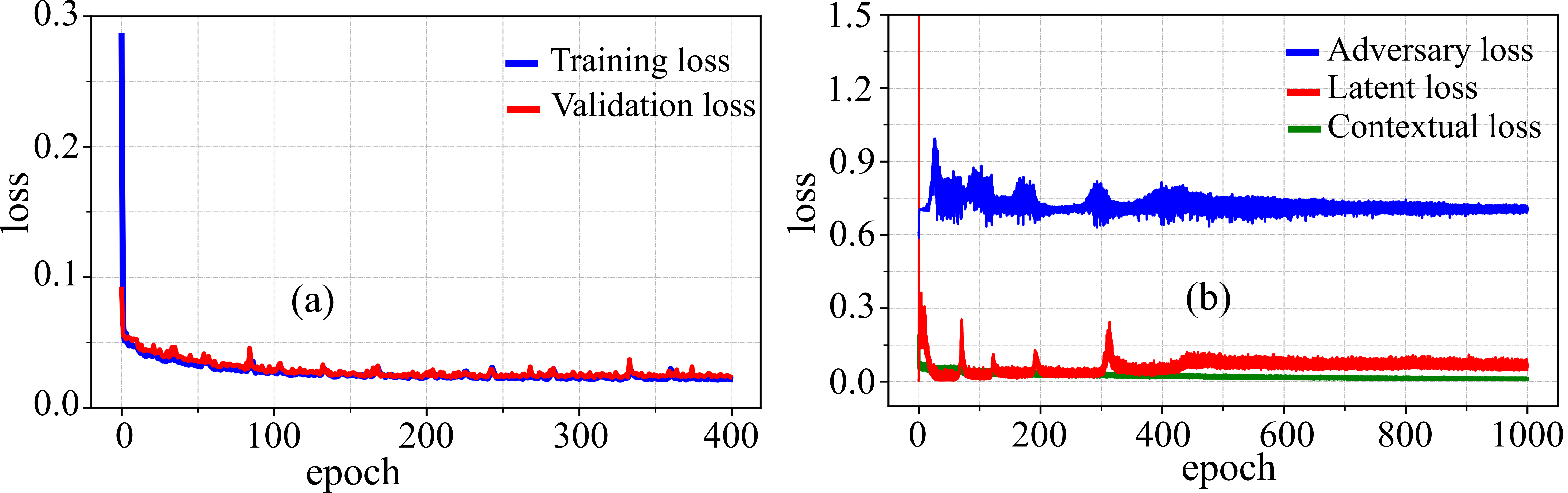}
    \caption{Example of model loss versus training epochs for (a) Autoencoder based anomaly detection and (b) GANomaly model.}
    \label{fig:loss}
\end{figure}

First of all, regarding the AUE model, it is important to ensure that it is not overfitting. In addition, for the GANomaly model the convergence is of great importance since it is adversarially trained and hence it is potentially unstable. To assure these things are not happening in our work, Fig.~\ref{fig:loss} is presented.  
From Fig.~\ref{fig:loss} it is clear that the AUE is not overfitting. Moreover, the GANomaly loss functions reaches stability after a number of training epochs.

Table~\ref{tab:result_total_metrics} shows the results of all the models in terms of the evaluation metrics. The results are based on the mean value and one standard deviation of five runs for Autoencoder, GANomaly, and CNN-BiLSTM and ten runs for other models. From this table we can see that in terms of the F1-score and Balanced accuracy, the proposed GANomaly model outperforms all the other models. In addition, the unsupervised models perform  better than the proposed supervised models. It should be pointed that as it was expected, the supervised deep classifier (CNN-BiLSTM + Attention) does not perform well. More specifically, although it shows a high recall value (even with high standard deviation), it has a very poor performance in precision.  

\begin{figure}[htbp!]
    \centering
    \includegraphics[width=1\linewidth]{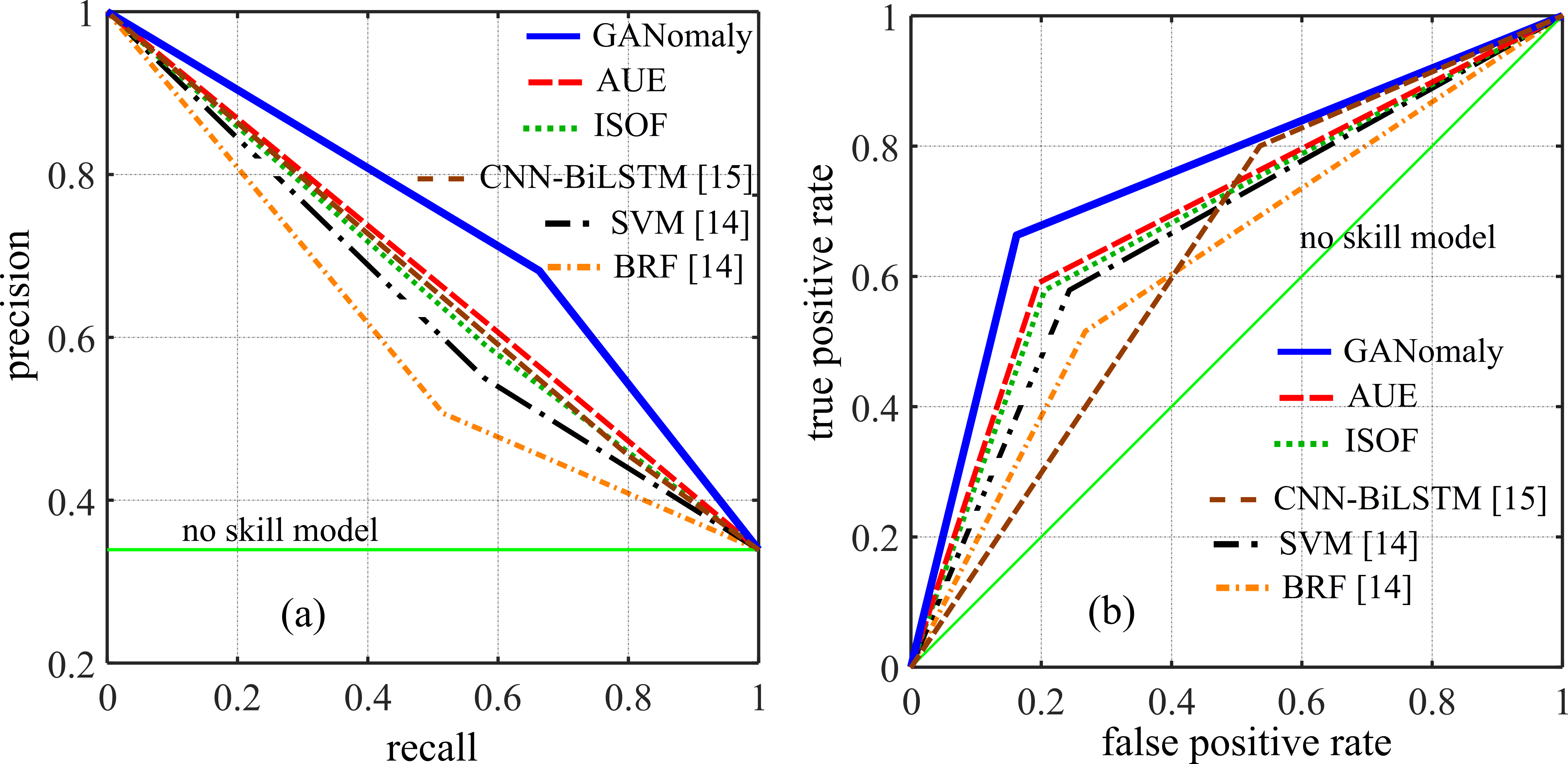}
    \caption{Comparison of the proposed models applied to test data samples in terms of (a) precision-recall curve and (b) receiver operating characteristic curve. A no skill model is a dummy classifier that assigns a random class label regardless of its input. Thus, it has a false positive rate equals to its true positive rate while its precision-recal curve is a horizontal line as the proportion of positive (abnormal) samples to the total.}
    \label{fig:prec_roc}
\end{figure}

As an another comparison of the performances of these models, the receiver operating characteristics (roc) curves and precision-recall curves are represented in Fig.~\ref{fig:prec_roc}. From Fig.~\ref{fig:prec_roc} (a) it can be seen that the modified GANomly has the highest area under the curve (AUC). In roc curve, in addition to having higher AUC, a higher TPR with small FPR is of our interest. Looking at Fig.~\ref{fig:prec_roc} (b), the GANomaly has the best roc curve and so the best performance among all the models. 

\begin{table*}[htbp]
	\centering
	\caption{Comparison of the performance of different models based on evaluation metrics applied to test data samples. For each model, the average and one standard deviation ( of five runs for Autoencoder, GANomaly, and CNN-BiLSTM and ten runs for other models) are shown.}
	\begin{adjustbox}{width=0.95\textwidth}
		\begin{tabular}{c l c c c c}
			\toprule
            \textbf{Approach} & \textbf{Model} & \textbf{F1-score} &\textbf{Balanced accuracy}&\textbf{Precision}& \textbf{Recall}\\
            \midrule[0.1pt]
            
            \multirow{3}{*}{\makecell{Supervised models\\ (Classification)} }
            &\text{CNN-BiLSTM + Attention, DWT~\cite{liu2021attention}}& $0.565\pm0.104$& $0.632\pm0.053$& $0.455\pm0.082$& $0.800\pm0.146$\\ \cmidrule(lr){2-6}&\text{Balanced Random Forest~\cite{ajirak2022boost}}& $0.622\pm0.042$ & $0.624\pm0.039$ & $0.507\pm0.078 $& $0.516\pm0.061$\\ \cmidrule(lr){2-6}
            &\text{Support Vector Machine~\cite{ajirak2022boost}}& $0.666\pm0.00$& $0.668\pm0.00$& $0.550\pm0.00$& $0.579\pm0.00$\\
            
            \midrule[0.1pt]
            
            \multirow{3}{*}{\makecell{Unsupervised/semi-supervised\\ models}}&\text{Isolation Forest}& $0.687\pm0.045$&$ 0.687\pm0.046$&$0.591\pm0.058$ &$ 0.579\pm0.085$\\ \cmidrule(lr){2-6}
            &\text{Autoencoder}& $0.699\pm0.018$& $0.697\pm0.013$& $0.613\pm0.043$& $0.589\pm0.021$\\ \cmidrule(lr){2-6}
            &\text{GANomaly}&$0.752\pm0.011$& $0.750\pm0.001$& $0.682\pm0.041$& $0.663\pm0.042$ \\

			\bottomrule
		\end{tabular}
	\end{adjustbox}
	\label{tab:result_total_metrics}
\end{table*}

\begin{figure}[htbp!]
    \centering
    \includegraphics[width=0.99\linewidth]{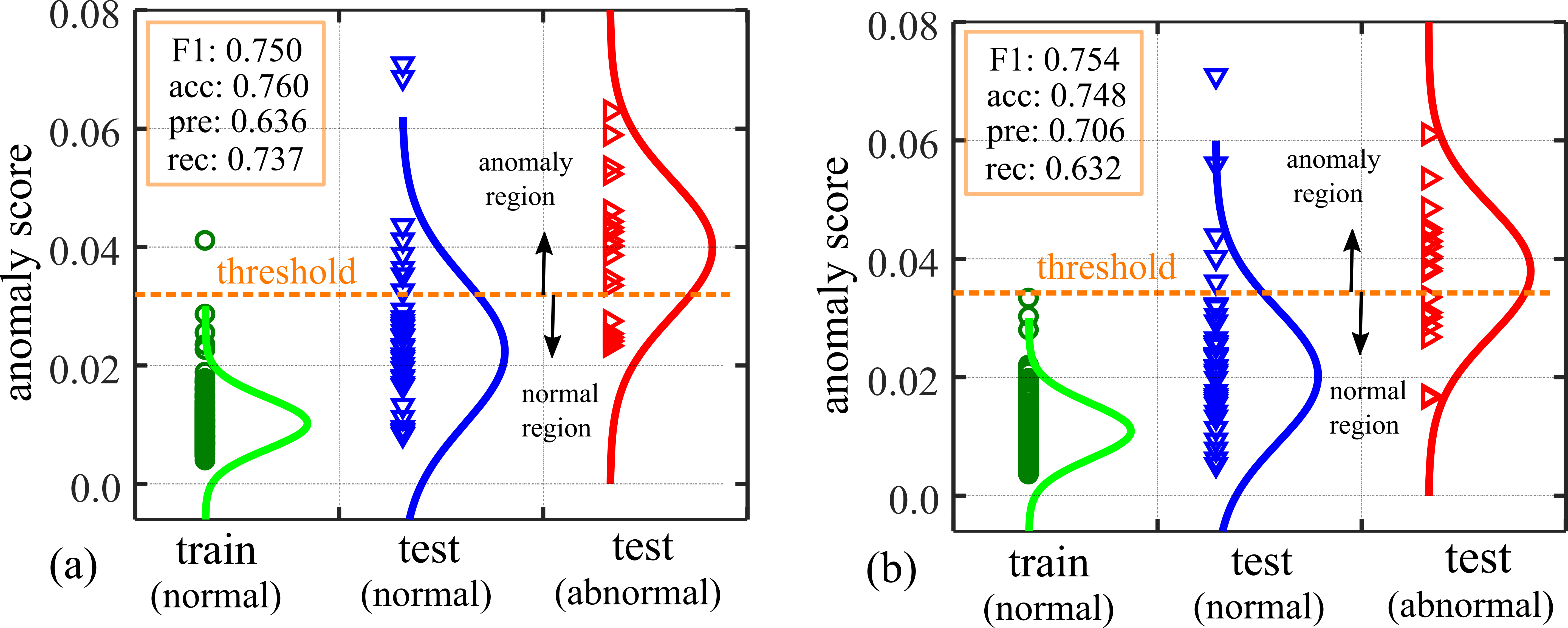}
    \caption{Two examples of the anomaly score values of train and test data samples based on the GANomaly model. The threshold value is set to have the maximum F1-score of anomaly detection for validation data samples.}
    \label{fig:score_dist}
\end{figure}

To get a better insight about the modified GANomaly model and observe the anomaly scores returned by this model, the anomaly score distributions of training data samples and test data samples are visualized in Fig.~\ref{fig:score_dist} for two runs of GANomaly
(with the same data samples but different random seeds used in initializing the learning parameters of model). It is worth to be mentioned again that the value of the threshold score is set to have the best F1-score on the validation dataset. Moreover, it should be noted that the histogram of the scores have a bell shape; that is why in Fig.~\ref{fig:score_dist} a Gaussian distribution is fitted to the scores. However, we did not show the histograms for the sake of having a clear visualization.

From Fig.~\ref{fig:score_dist} we see that the GANomaly model assigned smaller anomaly scores to the normal class compared to abnormal data samples. While in the example (a) the model can detect more abnormal samples (high recall) and in example (b) it has better precision in detecting abnormal samples, for both examples the model performs almost the same in terms of F1-score. 

In light of the results, we can perceive several advantages and disadvantages linked to the use of a deep unsupervised anomaly detection model. First of all, it is important to remember that the CTU-UHB database does not represent the entire population of pregnant women. For example, preterm deliveries (i.e. gestational age $<$ 36 weeks) have been excluded from the database in order to obtain a homogeneous database \cite{chudavcek2014open}. In a real world application, the model risks assimilating a preterm pregnancy to an anomaly since it has not been trained with similar samples. Therefore, the model should be confronted to a greater variety of samples, e.g. using the same dataset as Petrozziello et al. \cite{petrozziello2019mcnn}.
Although it is important to use more data, our model only requires non-pathological samples for training, which is a great advantage. Indeed, classification-based methods need both classes in quantity, including pathological samples that are the most difficult to collect since they are statistically rarer. In future works, it is therefore easier to increment the existing database with new data using our anomaly detection model. 


Other extensions to this work can be considered. The first one is to jointly analyze the UC and FHR signals where it is possible (in terms of the quality of the available signals). This might improve the performance of the proposed deep unsupervised model. The second one could be studying other approaches for labeling the CTG signals, e.g using information coming from the umbilical blood sample or considering other pH values such as 7.05 that refers to pathological cases. Finally, in dealing with missing values, a more sophisticated approach (other than linear interpolation), e.g. FHR prediction using recurrent neural networks, might improve the performance. Those extensions could benefit the physicians in the interpretation of CTG by giving them more insights on the AI point of view.



Finally, it is also interesting to highlight the interest of this technology for the medical world. A model of anomaly detection could help doctors in their daily work to notice pathological analyses more quickly. In our study, we were particularly interested in childbirth but the principle could be extended to other moments of pregnancy. Indeed, fetal monitoring is used throughout the pregnancy to check the well-being of the fetus and the doctor is not always present to make a diagnosis on the spot. The use of an anomaly detection program could therefore also be relevant outside of pregnancy to save time. 

\section{Conclusion} \label{sec:conclusion}
Fetal health monitoring based on artificial intelligence has received a great attention in the recent years. Most of the proposed models that are based on the supervised learning lack generalization to test (unseen) data samples. Those studies who used unsupervised approaches either reported inefficiency of their models or their work was limited to a certain number of CTG signals with high confident (normal/abnormal) labels. In this work, we adopted using deep unsupervised models trained in a semi-supervised manner. The GANomaly framework with a modified loss function for capturing the underlying distribution of data samples, was used as our main model and was applied to all the CTG samples in the CTU-UHB dataset. Experimental comparison of the proposed modified GANomaly with the other unsupervised models, such as Isolation Forest and Autoencoder, or the supervised models used in the recent studies, including SVM, Balanced Random Forest, and CNN-BiLSTM with attention network showed the superiority of the deep unsupervised approaches in detecting suspicious CTGs. More specifically, the modified GANomaly showed the best performance (on the test data samples) among the proposed models in terms of the F1-score, balanced accuracy, precision-recall and roc curves.

\section*{Acknowledgement}
This work was supported by Hydro-Quebec, the Natural Sciences and Engineering Research Council of Canada, and McGill University in the framework of the NSERC/Hydro-Quebec Industrial Research Chair in Interactive Information Infrastructure for the Power Grid (IRCPJ406021-14).

\bibliographystyle{ieeetr}
\bibliography{HREF}

\end{document}